\documentclass[reprint,aps,prd,amsmath,showpacs]{revtex4-1}
\usepackage{graphicx}
\usepackage[linkcolor=blue,citecolor=blue]{hyperref}
\usepackage{bm}
\usepackage{float}
\usepackage{amssymb}
\usepackage{latexsym}
\usepackage{scrextend}
\hypersetup{colorlinks=true}

\newcommand{\mm}{\mu\bar{\mu}}

\begin{document}
\title{Hadronic Vacuum Polarization in True Muonium}

\author{Henry Lamm}
\email{hlammiv@asu.edu}
\affiliation{Department of Physics, Arizona State University, Tempe, AZ 85287-1504, USA}
\date{\today}

%%%%%%%%%%%%%%%%%%%%%%%%%%%%%%%%%%%%%%%%%%%%%%%%%%%%%%%
\begin{abstract}
The leading-order hadronic vacuum polarization contribution to the hyperfine splitting of true muonium is reevaluated in two ways.  The first considers a more complex pionic form factor and better estimates of the perturbative QCD contributions.  The second, more accurate method directly integrates the Drell ratio $R(s)$ to obtain $C_{1,\rm hvp}=-0.0489(3)$.  This corresponds to an energy shift in the hyperfine splitting of $\Delta E^\mu_{hfs,\rm hvp}=276196(51)$ MHz.
\end{abstract}
%%%%%%%%%%%%%%%%%%%%%%%%%%%%%%%%%%%%%%%%%%%%%%%%%%
\maketitle
\section{Introduction}
True muonium is the yet unidentified $(\mm)$ bound state.  The bound states have lifetimes between ps to ns~\cite{Brodsky:2009gx}.  QED dominates the characteristics of true muonium, while QCD effects appear at $\mathcal{O}(m_\mu\alpha^5)$~\cite{Jentschura:1997tv,Jentschura:1997ma}. Electroweak effects appear at $\mathcal{O}(m_\mu\alpha^7)$~\cite{PhysRevD.91.073008}.  Measurements of Lamb shift, $1s-2s$ splitting, and the hyperfine splitting (hfs) will occur in the future.  These experiments are motivated by the existing discrepancies in muon physics~\cite{PhysRevD.73.072003,Antognini:1900ns,Aaij:2014ora,Aaij:2015yra,Pohl1:2016xoo}.  Numerous new physics models have been suggested to explain these discrepancies~\cite{TuckerSmith:2010ra,Jaeckel:2010xx,Batell:2011qq,Barger:2011mt,Karshenboim:2010cm,Karshenboim:2010cg,Karshenboim:2010cj,Karshenboim:2010ck,Karshenboim:2011dx,Karshenboim:2014tka,Carlson:2012pc,Carlson:2015poa,Izaguirre:2014cza,Kopp:2014tsa,Martens:2016zzx,Liu:2016qwd,Onofrio:2013fea,Wang:2013fma,Gomes:2014kaa,Brax:2014vva,Lamm:2015gka,Lamm:2016jim}.  True muonium can produce competitive constraints on most models if standard model predictions are known to the 100 MHz level, corresponding to $\mathcal{O}(m_\mu\alpha^7)$.  

Beyond new physics, a further motivation for considering true muonium comes from the anomalous magnetic moment of the muon ($a_\mu$).  There exists a discrepancy between the measurement at BNL and theory, $\Delta a_\mu=a_{\mu,exp}-a_{\mu,th}=288(80)\times10^{-11}$~\cite{PhysRevD.73.072003,Aoyama:2014sxa}.  Hadronic contributions dominate the theoretical uncertainty, and hadronic vacuum polarization (hvp) is the largest term.  One way to reduce the theoretical uncertainty would be consistency checks from other systems.  By its particle/antiparticle nature, the annihilation channel contributes to true muonium, leading to an enhancement of hvp contributions to the hfs.  These contributions are measurable in true muonium unlike positronium were they are mass-suppressed.

The theoretical expression for the hfs corrections to true muonium from QED can be written
\begin{align}
 \Delta E_{\rm hfs}=m_\mu\alpha^4\bigg[&C_0+C_1\frac{\alpha}{\pi}+C_{21}\alpha^2\ln\left(\frac{1}{\alpha}\right)+C_{20}\left(\frac{\alpha}{\pi}\right)^2\nonumber\\
 &+C_{32}\frac{\alpha^3}{\pi}\ln^2\left(\frac{1}{\alpha}\right)+C_{31}\frac{\alpha^3}{\pi}\ln\left(\frac{1}{\alpha}\right)\nonumber\\
 &+C_{30}\left(\frac{\alpha}{\pi}\right)^3+\cdots\bigg],
\end{align}
where $C_{ij}$ indicate the coefficient of the term proportional to $(\alpha)^i\ln^j(1/\alpha)$.  All dependence of the hfs to mass scales other than $m_\mu$ is in the $C_{ij}$.  The coefficients of single flavor QED bound states, used in positronium, are known up to $\mathcal{O}(m_e\alpha^6)$ and some partial results for $\mathcal{O}(m_e\alpha^7)$(For an updated review of the coefficients see \cite{Adkins:2014dva,PhysRevA.94.032507}).  The exchange $m_e\rightarrow m_\mu$ translates these results to true muonium.

True muonium has extra contributions that must be considered.  The lighter electron allows for large loop contributions.  The relative smallness of $m_\tau/m_\mu\approx 17$ and $m_{\pi}/m_{\mu}\approx1.3$ produce contributions to true muonium much larger than analogous contributions to positronium.  Of these true muonium specific contributions, which we denote by $C_{ij}^\mu$, only a few terms are known.  The $\mathcal{O}(m_\mu\alpha^5)$ contributions from electron loops were found to be $C_{1,e}^\mu=1.684$~\cite{Jentschura:1997tv}.  The $\mathcal{O}(m_\mu\alpha^6)$ contribution from leptonic loops to the two-photon annihilation channel $C^\mu_{20,2\gamma}=-2.031092873$ was recently computed exactly~\cite{PhysRevA.94.032507}, and the electron loop in three-photon annihilation at $\mathcal{O}(m_\mu\alpha^7)$ is $C_{30,3\gamma}^\mu=-5.86510(20)$~\cite{Adkins:2015jia}.  For a $\mathcal{O}(m_\mu\alpha^7)$ prediction of the hfs, contributions from $Z$-bosons must be considered~\cite{PhysRevD.91.073008}.

The hadronic vacuum polarization contributes at $\mathcal{O}(m_\mu\alpha^5)$ and was previously calculated to be $C_{1,\rm hvp}=-0.047(5)$ in ~\cite{Jentschura:1997tv} where the error is an estimate of the model-dependence.  We will refer to this result as JSIK throughout, after the authors of that paper.  This result mixed a Gounaris-Sakurai form factor for the $\pi$ and $\rho$ contributions, a simple pole approximation for the $\omega$ and $\phi$, and a two-constant perturbative contribution above 1 GeV.

Together, these contributions predict the hfs of true muonium to be $\Delta E_{\rm hfs}^{1s}=42329730(800)(700)$ MHz where the first, dominant, uncertainty is from the hadronic model-dependence and the second is an estimate of uncalculated $\mathcal{O}(m_\mu\alpha^6)$ contributions.  The goal of this work is to recalculate $C_{1,\rm hvp}$ such that we can both reduce the model dependence and better estimate the uncertainty.

The hadronic vacuum polarization contribution is given by 
\begin{align}
\label{eq:energy}
 \Delta E_{1,\rm hvp}=&\left[m_\mu^2\int_{4m_\pi^2}^{\infty}\mathrm{d}s\frac{\rho(s)}{4m^2_{\mu}-s}\right]\frac{m_\mu\alpha^5}{n^3\pi}\nonumber\\
 =&C_{1,\rm hvp}\frac{m_\mu\alpha^5}{n^3\pi}
\end{align}
where $\rho(s)$ is the spectral function that must be specified.
\section{JSIK Calculation}
The calculation of $C^{\rm JSIK}_{1,\rm hvp}$ in~\cite{Jentschura:1997tv} is given by the sum of four terms,
\begin{equation}
 C^{\rm JSIK}_{1,\rm hvp}=C_{1,\pi}+C_{1,\omega}+C_{1,\phi}+C_{1,>}
\end{equation}
where $C_{1,\pi}$ is the contribution from the pion form factor, $C_{1,\omega}$ and $C_{1,\phi}$ are the simple-pole terms for these two meson, and $C_{1,>}$ is the contribution from the regime above 1 GeV were perturbative QCD was applied.

The main contribution is from the pionic loop, where the result is given by~\cite{Sapirstein:1983xr,Jentschura:1997tv}
\begin{equation}
\rho(s) = \frac{(s - 4\,m_{\pi}^2)^{3/2}}{12\,s^{5/2}}\,
|F_{\pi}(s)|^2 \,.
\end{equation}
JSIK chose to use the simple Gounaris-Sakurai form factor~\cite{Gounaris:1968mw}.  For brevity, this choice of $F_{\pi}(s)$ is often written as
\begin{equation}  \label{eq:fpi}
F_{\pi}(s)=F_{\rho,\rm GS}(s) = \frac{N}{D_1 + D_2 - i\,D_3}\,.
\end{equation}
In this decomposition, $N$, $D_1$, $D_2$ and $D_3$ are given by
\begin{align}
N = m_\rho^2 + d m_\rho \, \Gamma_\rho\,,&\phantom{xxx}D_1 = m_\rho^2 - s,\nonumber\\
D_2 = \Gamma_\rho \frac{m_\rho^2}{k_\rho^3}
\bigg[k(s)^2(h(s) &- h_\rho) + k_\rho^2 h'_\rho
(m_\rho^2 - s) \bigg],\nonumber\\
\quad D_3 = \frac{m_\rho^2\Gamma_\rho}{\sqrt{s}} & 
\left(\frac{k(s)}{k_\rho}\right)^3 \,,
\end{align}
with the parameter $d$ defined via
\begin{equation}
d = \frac{3}{\pi} \frac{m_\pi^2}{k_\rho^2} 
\ln\frac{m_\rho + 2 \, k_\rho}{2\,m_\pi} + 
\frac{m_\rho}{2\,\pi\,k_\rho} - 
\frac{m_\pi^2\, m_\rho}{\pi\,k_\rho^3} \approx 0.48.
\end{equation}
 
The functions $k(s)$ and $h(s)$ are defined as
\begin{equation}
k(s)=\frac{1}{2}\sqrt{s - 4m_\pi^2}, \quad h(s) = \frac{2}{\pi} \,
\frac{k(s)}{\sqrt{s}} \, 
\ln \left(\frac{\sqrt{s} + 2\, k(s)}{2\,m_\pi}\right)\,.
\end{equation}
Where $h'$ denoted the derivative of $h(s)$ with respect to $s$ and the subscript $\rho$ indicated evaluation of the function at $m^2_\rho$.  
In this form factor, only the contributions for the $\rho$ mesons are included.  The physical values used by JSIK were $\Gamma_\rho = 150.7(1.2) \, {\rm MeV},$ and $m_\rho = 768.5(6) \, {\rm MeV}$.  Integrating these expressions, the value for the pionic loop was found to be $C_{1,\pi}=-0.032$.

To include other meson resonances, a simple pole approximation is taken.  The spectral function contribution from a vector meson is given by $\rho(s)=4\pi^2/f_V^2\delta(s-m_V^2)$~\cite{Bauer:1977iq} where $f_V$ are coupling constants.  These were estimated in~\cite{Bauer:1977iq} to be $f_\omega^2/4\pi=18(2)$ and $f_\phi^2/4\pi=11(2)$.  The masses of the vector mesons are $m_\omega=782.71(8)$ MeV and $m_\phi=1019.461(19)$ MeV. JSIK obtained $C_{1,\omega}=-0.004$ and $C_{1,\phi}=-0.003.$

\begin{figure*}
 \includegraphics[width=\linewidth]{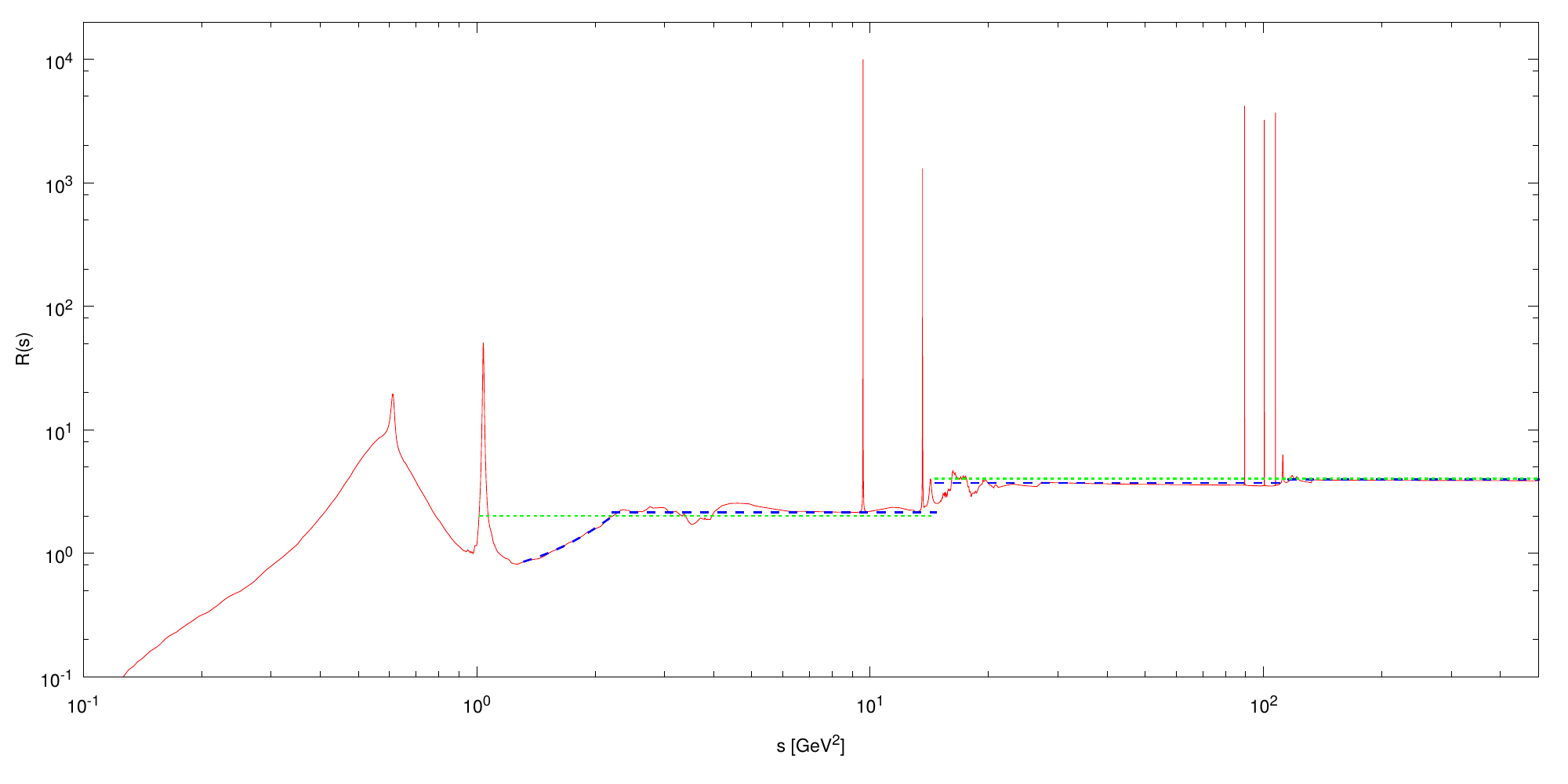}
 \caption{\label{fig:rs}$R(s)$ vs. $s$.  The solid line indicates the experimental results from the compilation used by alphaQED~\cite{Jegerlehner:2003qp,Jegerlehner:2003rx,Jegerlehner:2003ip,Jegerlehner:2015stw} and rhad~\cite{HARLANDER2003244}. The dotted line are the estimates used in the JSIK calculations of $C_{1,>}$~\cite{Jentschura:1997tv}, and the dashed line indicates the estimates of this work.}
\end{figure*}

The final contribution, $C_{1,>}$ was obtained by applying the relation between the spectral function and the Drell ratio,
\begin{equation}
\label{eq:rhors}
 \rho(s)=\frac{R}{3s}, \text{ where } R=\frac{\sigma(e^+e^-\rightarrow h)}{\sigma(e^+e^-\rightarrow\mu^+\mu^-)}.
\end{equation}
In perturbative QCD, $R$ at leading order is given by $R_{LO}=N_c\sum q_i^2$ where $N_c$ is the number of colors and $q_i$ is the charge of quark $i$.  Below the $c$ threshold at $\approx4$ GeV, $R_{LO}=2$.  Between 4 GeV and 10 GeV, the $c$ quark goes on shell and $R_{LO}=10/3$.  Above 10 GeV the $b$ quark changes the ratio to $R_{LO}=11/3$.  At present, perturbative calculations exist up to $\mathcal{O}(\alpha_s^4)$ that better account for the experimental results.  JSIK estimated from the experimental results in~\cite{Barnett:1996hr} that $R_{\rm 2 GeV<s<4 GeV}\approx2$ and $R_{s>4 GeV}\approx4$ (See Fig.~\ref{fig:rs}).  With these values, they obtained $C_{1,>}=-0.008.$

Putting all of these together, and including a 11\% estimate of the model-dependent uncertainties, their final result was $C_{1,\rm hvp}^{\rm JSIK}=-0.047(5)$

\section{Investigating the Pieces}
One way to reduce the uncertainty in $C_{1,\rm hvp}$ would be to improve the calculations of the pieces of the JSIK value.  Since that work, improvements in experimental measurements of the pion form factor have lead to the development of an improved Gounaris-Sakurai parameterization that more properly accounts for the $\rho-\omega$ mixing as well as the $\rho'$ and $\rho''$ states.  $C_{1,>}$ can also be better estimated by computing the numerical averages of $R(s)$ in the regimes and accounting for non-constant terms.
\subsection{Improved Gounaris-Sakurai Parameterization}
Instead of the simple Gounaris-Sakurai form factor, a more complex form exists that features two improvements.  The improved form~\cite{Schael:2005am} includes $\rho-\omega$ mixing and the $\rho'$ and $\rho''$ resonances.  The form factor is given by
\begin{align}
 F_{\pi,\rm IGS}=\frac{1}{1+\beta+\gamma}\big[F_{\rho,\rm GS}&(s)\left(1+\delta \frac{s}{m^2_\omega}F_{\omega,\rm BW}(s)\right)\nonumber\\&+\beta F_{\rho',\rm GS}(s)+\gamma F_{\rho'',\rm GS}(s)\big]
\end{align}
where $F_{i,\rm GS}(s)$ are given by Eq.(\ref{eq:fpi}) with the additional masses and decay constants: $m_{\rho'}=1409(12)$ MeV, $\Gamma_{\rho'}=501(37)$ MeV, $m_{\rho''}=1740(21)$ MeV, $\Gamma_{\rho''}=235(1)$ MeV, and $\Gamma_\omega=8.68$ MeV~\cite{Schael:2005am}.  Further the parameters $\delta=2.03(10)e^{0.2269(401)i}$, $\beta=-0.166(6)$, and $\gamma=0.071(6)$ determine the mixing and relative strengths~\cite{Schael:2005am}.  For the $\omega$ meson, a Breit-Wigner form factor is used
\begin{equation}
 F_{\omega,\rm BW}(s)=\frac{m^2_{\omega}}{m^2-s+i\Gamma_{\omega}m_\omega}.
\end{equation}
Integrating, we compute a coefficient $C_{1,\rm IGS}=-0.0377(5)$ which should include the same physics as $C_{1,\pi}+C_{1,\omega}$ as the well previously uncalculated higher-order terms from $\rho',\rho''$.  The error on $C_{1,\rm IGS}$ is estimated from parameter variation.

\subsection{Perturbative QCD Regime} 
We use the data for R(s) compiled for the software packages alphaQED~\cite{Jegerlehner:2003qp,Jegerlehner:2003rx,Jegerlehner:2003ip,Jegerlehner:2015stw} and rhad~\cite{HARLANDER2003244}.  These packages may be found at~\cite{alphaqed}.  It can be seen in Fig.~\ref{fig:rs} that the JSIK value of $C_{1,>}$ can be improved. For $s\approx1$ GeV$^2$ to $s\approx2$ GeV$^2$, the JSIK estimate overestimates the contribution, and ignores the $s$-dependence.  We find that in the region $s=[1.2,2.3]$ GeV$^2$ is well fit to $R(s)=0.0895(9)s^{3.43(11)}+0.63(2)$.  Above this, we take the $R$ to be a constant fit to the average value without resonances.  Between $s=2.3$ GeV$^2$ and the $s=16$ GeV$^2$ $R\approx2.15(1)$.  In the region $s=[16,120]$ GeV$^2$, we find $R\approx3.71(1)$ and above this we take $R\approx3.95(1)$.  Together, these choices give a value of $C_{1,>}=-0.00574(4)$

Adding our values of $C_{1,\rm IGS}$ and $C_{1,>}$ to the previously obtained value of $C_{1,\phi}$, our final results for the improved piecewise coefficient is $C^{\rm imp}_{1,\rm hvp}=-0.0467(5)$.  This value represents an improvement on the JSIK value, but we note that it still has a large model-dependence which is difficult to estimate, and doesn't fully encapsulate the effect of resonances.

\section{Numerical Integration of $R(s)$}
Another method to obtain $C_{1,\rm hvp}$ is numerically integrating $R(s)$.  This method has negligible model-dependence and theoretical uncertainties.  We numerically integrate the full $R(s)$ data from Ref.~\cite{HARLANDER2003244,Jegerlehner:2003qp,Jegerlehner:2003rx,Jegerlehner:2003ip,Jegerlehner:2015stw,alphaqed} seen in Fig.~\ref{fig:rs} using Eq.(\ref{eq:energy}) and Eq.(\ref{eq:rhors}).  Without interpolation, the values of $R(s)$ tend to be overestimated due to the step behavior from binning in the data, especially around resonances.  To avoid this, we interpolate the data with $n-$order polynomials before numerical integration.  The results for $C_{1,\rm hvp}$ are found in Table~\ref{tab:rs}.  We average these values to yield our final result, $C^R_{1,\rm hvp}=-0.0489(3)$.  The error is the standard deviation of the interpolated results.

By replacement of $m_\mu\rightarrow m_e$ we can also compute the correction to positronium.  We find that value to be $C^{R,e}_{1,\rm hvp}=-1.030(6)\times10^{-6}$, which is too small to be relevant in the near-future.
\begin{table}
 \caption{\label{tab:rs}$C^R_{1,\rm hvp}$ from directly integrating $R(s)$ for both true muonium and positronium.  The order indicates the polynomial order used to fit the experimental data.}
 \begin{center}
 \begin{tabular}{c c c}
 \hline\hline
  Order&$C^{R,\mu}_{1,\rm hvp}$&$C^{R,e}_{1,\rm hvp}$\\
  \hline
  0	& -0.05001 &-1.051$\times10^{-6}$\\
  1	& -0.04879 &-1.028$\times10^{-6}$\\
  2	& -0.04874 &-1.027$\times10^{-6}$\\
  3	& -0.04869 &-1.026$\times10^{-6}$\\
  4	& -0.04846 &-1.021$\times10^{-6}$\\
  \hline
  Avg.	& -0.0489(3)&-1.030(6)$\times10^{-6}$\\
  \hline\hline
 \end{tabular}
\end{center}
\end{table}
\section{Summary and Conclusion}
In this work, we have recomputed the coefficient $C_{1,\rm hvp}$ in two ways.  The first was improving upon the work of~\cite{Jentschura:1997tv} through the use of a more complex pionic form factor and better modeling of the perturbative regime. The final calculation in this technique was $C^{\rm imp}_{1,\rm hvp}=-0.0467(5)$, where the error was only estimated by parameter variation and would miss systematic errors.  While more precisely accounting for some of the features of the full spectral function, it has some drawbacks.  It treats the $\phi$ meson as a simple pole, which will underestimate its contribution.  Further, the treatment of all physics above 1 GeV$^2$ by the perturbative background neglects resonances and other features.  

In order to avoid these problems, we computed $C_{1,\rm hvp}$ directly from experimental $R(s)$.  To account for the binning of the data, we used $n-$order polynomial interpolators.  The final value for this method was $C^R_{1,\rm hvp}=-0.0489(3)$.  This value is larger than both JSIK and our improved method by more than estimated error of $C^{\rm imp}_{1,\rm hvp}$.  We attribute this to the inadequate treatment of resonances in the these methods.  Further, the model-dependence inherent in fitting the form factor has been avoided in this method.  Therefore, we take this as our final value for calculations of the hfs.

 With this contribution found, we can reevaluate the prediction for hfs.  Our result reduces the hadronic error estimate of JSIK\cite{Jentschura:1997tv} from $800$ MHz to $51$ MHz.  The current value is $\Delta E^{1s}_{\rm hfs}=42329355(51)_{\rm had}(700)_{\rm miss}\text{ MHz}$,
where the first uncertainty is from hadronic contribution, and the second an estimate of missing $\mathcal{O}(m_\mu\alpha^6)$ terms.  With this reduction in hadronic uncertainty, the missing QED corrections now dominates and is the only remaining step to obtaining $\mathcal{O}(100$ MHz) predictions for use in new physics searches.

\begin{acknowledgments}
HL would like to thank E Lee for her assistance in locating an error in the numerical integration.  HL is supported by the National Science Foundation under Grant Nos. 
PHY-1068286 and PHY-1403891.
\end{acknowledgments}

%\bibliographystyle{apsrev4-1}
%\input{Eloop.bbl}
%\bibliography{Eloop}
\bibliography{/home/hlamm/wise}
\end{document}